\documentstyle[aps,prb,array,
multicol,bezier
]{revtex}
\begin{document}
\title{
   The Kondo state in quantum point contacts and the local moment in
semiconductor quantum dots 
   - two sides of the same phenomenon.
}
\author{ Daniel L. Miller }
\address{ Intel Electronics, P.O.Box 3173\\
          Jerusalem 91031, Israel \\
          Daniel.Miller@Intel.Com 
}
\date{\today}
\maketitle
\begin{abstract}
   This is a three step work: i) we explain why quantum point contacts are
similar 
to ballistic quantum dots; ii) we introduce the virtual Kondo state in both
systems;
iii-1st) this state explains 0.7 structure in point contacts; 
iii-2nd) formation of the local moment on this state is described by the
nearly symmetric Anderson model, we 
solve it for finite size system having in mind quantum dots. We found one
large level spacing 
$\Delta^\ast \propto (U\Gamma)^{1/2}\gg \Delta$, where $U$ is the charging
energy of the virtual state, 
$\Gamma$ is the spectral width of this state and $\Delta$ is the mean level
spacing of whole system. 
The theory explains periodicity of abnormal level spacing vs gate potential.
The theory is in agreement 
with many experiments. 
\end{abstract}

\begin{multicols}{2}
\narrowtext


\section{ Why quantum point contacts are similar to chaotic quantum dots. }

In this work we develop common view on two types of experiments, first,
differential conductivity
of quantum point contacts\cite{Wees-1988,Thomas-1996,Cronenwett-2002}, and
second, fluctuations of level spacing 
in semiconductor quantum
dots\cite{Sivan-aug96,Zhitenev-sep97,Patel-may98,Patel-aug98,Simmel-jan99,Lu
cher-mar2001}. 
The challenge in the former 
case is to understand the plato of differential conductance $dI/dV \approx
0.7(2e^2/h)$, which indicates presence
of the Kondo state\cite{Cronenwett-2002}. The Kondo state can appear from a
magnetic impurity, 
as in the case of metallic
point contacts\cite{Ralph-Buhrman-1994}, however the point contact was made
by electrodes on top of 
clean 2d electron gas\cite{Wees-1988,Thomas-1996,Cronenwett-2002}.

The level spacing fluctuation in the latter case indicates presence of a
local moment inside the 
dot\cite{Berkovits-1998}. There are two possibilities: either magnetic
moment is formed on the 
localized Kondo state or the magnetic moment is formed due to the Stoner
instability of electron 
system\cite{Andreev-Kamenev-1998,Brower-Oreg-Halperin-99,Kurland-2000,Barang
er-Ullmo-Glazman}.
However, the clean semiconductor quantum dots do not have magnetic
impurities and formation of the Kondo state
is unrealistic. The Stoner instability occurs under very restrictive
condition met only in 
some experiments\cite{Jacquod-Stone-2001,Folk-2000}.

The anticipated Kondo state can be found in both systems, quantum point
contact and semiconductor quantum dot,
as a virtual state bouncing along shortest axis of the system. This state is
marked by dashed line in 
Fig. 1a for contact geometry and in Fig. 1b for real potential of
rectangular 
quantum dot\cite{Stopa-sep97,Stopa-1999}. We assume that the system is
ballistic, $l\gtrsim l_p$, where $l$ is the mean
free path and $l_p$ is the length of the virtual state. This state is very
unstable indeed, the electron lifetime on such 
orbit $\hbar/\Gamma$ is of the order of the time of flight $l_p/v_F$, in
other words level uncertainty of such 
state is of the order of level spacing $\Gamma\sim  {\pi \hbar^2 k_F/ml_p}$,

where $k_F$ is the Fermi momentum and $m$ is the effective mass. This state
is important 
for two main reasons i) it has good coupling to all other ``chaotic'' wave
functions\cite{Berry-jul77,Heller-oct84}
ii) it cost extra energy to put two electrons on this state compare to all
other ``chaotic'' wave functions spread 
across whole system; the enhancement factor is approximately $\log(k_F
l_p)$.

\begin{figure*}
\unitlength=1mm
\special{em:linewidth 0.4pt}
\linethickness{0.4pt}
\begin{picture}(80.00,90.00)(0.0,-10.0)
\put(10.00,70.00){\line(1,0){20.00}}
\put(10.00,60.00){\line(1,0){20.00}}
\bezier{128}(30.00,70.00)(45.00,65.00)(30.00,60.00)
\put(80.00,70.00){\line(-1,0){20.00}}
\put(80.00,60.00){\line(-1,0){20.00}}
\bezier{128}(60.00,70.00)(45.00,66.00)(60.00,60.00)
\put(40.00,65.00){\line(1,0){2.00}}
\put(44.00,65.00){\line(1,0){2.00}}
\put(48.00,65.00){\line(1,0){2.00}}
\bezier{292}(10.00,40.00)(45.00,30.00)(80.00,40.00)
\bezier{292}(10.00,0.00)(45.00,10.00)(80.00,0.00)
\bezier{180}(10.00,40.00)(20.00,20.00)(10.00,0.00)
\bezier{180}(80.00,40.00)(70.00,20.00)(80.00,0.00)
\put(45.00,33.00){\line(0,-1){2.00}}
\put(45.00,29.00){\line(0,-1){2.00}}
\put(45.00,25.00){\line(0,-1){2.00}}
\put(45.00,21.00){\line(0,-1){2.00}}
\put(45.00,17.00){\line(0,-1){2.00}}
\put(45.00,13.00){\line(0,-1){2.00}}
\put(45.00,9.00){\line(0,-1){2.00}}
\put(5.00,70.00){\makebox(0,0)[ct]{(a)}}
\put(5.00,40.00){\makebox(0,0)[ct]{(b)}}
\end{picture}
\caption{ Local moment (Kondo state) is formed on unstable periodic
trajectory between tips of quantum point 
contact (a) or across quantum dot (b). This orbit traps electrons for some
time and have high charging energy.
}
\label{fig:qpc-qdot}
\end{figure*}
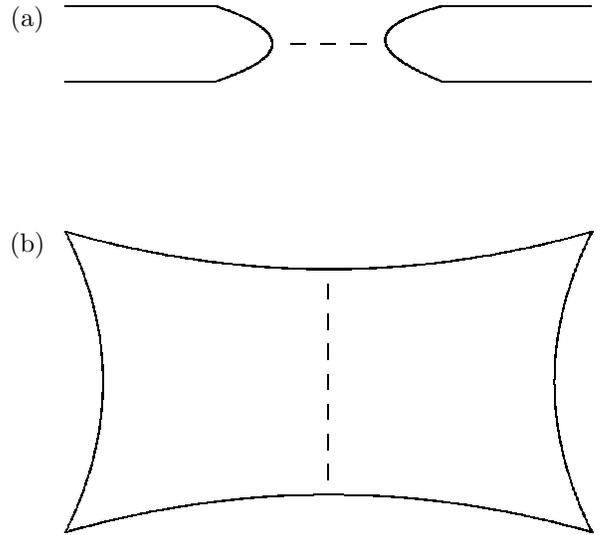

The sweep of the gate potential in both dots and contacts raise the Fermi
level; 
the singly occupied Kondo state floats on the Fermi
level\cite{Haldane-feb78,Stopa-nov96}. 
In this way conductance of the point contact at zero bias is given by 
resonant conductance of Kondo state. The level spacing of the quantum dot is
almost 
the same as of a non-interacting system. When the Fermi level reaches
``symmetric'' 
limit position, the virtual state can be doubly occupied. The system becomes
effectively non-interacting.
At this gate potential one observes kink of the level spacing in the quantum
dot.
In other words, the termodynamic limit of the symmetric Anderson model is
incompressible electron liquid,
and finite size system has just pseudogap.

In the next section we derive effective Hamiltonian by making use of Green
function technique and 
arrive at the Anderson model. The nearly symmetric limit is better to
explore 
by making use of the Wiegmann solution, that is done in Section III. We
solve Bethe equation with one energy level 
split out of the band, the ``gap'' energy is exactly the experimentally
observed kink in the level spacing. 
Section IV provides analysis of some experimental data and summary.


\section{ Hamiltonian of the virtual state. }

Single electron wave functions $\psi_j(\vec r)$ of a classically chaotic
system are inhomogeneous.  Consider an unstable periodic orbit $p$ in a
system with two degrees of freedom. Such an orbit contributes to the density
of states at the energy $\epsilon$ the sum\cite{Gutzwiller-mar71}
\begin{equation}
    {- T_p\over 2\pi\hbar}   \text{ Im }
    \sum_{m=1}^\infty
    {1\over \sinh(\Lambda_p m/2)}
    e^{im S_p(\epsilon+i0)/\hbar} \;,
\label{eq:trace.1}
\end{equation}
where $T_p={\partial S_p/\partial \epsilon}$ is the period of the orbit, 
$S_p(\epsilon)$ is the action of the orbit, $S_p(\epsilon)$ includes the
phase changes at the
conjugated points, $\Lambda_p>0$ is the stability exponent.

Let us assume that $p$ is the only one relatively stable $\Lambda_p \lesssim
1 $
periodic orbit in the system. In this particular case $p$ contributes
a sequence of Lorentzians\cite{Gutzwiller-mar71} to the density of states:
\begin{equation}
    \sum_n
    T_p(\epsilon) { \Lambda_p \hbar \over
   [ S_p(\epsilon) - 2\pi n \hbar ]^2 + (\Lambda_p \hbar /2)^2 }\;.
\label{eq:trace.3}
\end{equation}
Each term in the sum can be regarded as the spectral function of the
``scar''
state
\begin{equation}
   A_n(\epsilon) =
   {1\over \pi} {\Gamma_n \over ( \epsilon -  \epsilon_n)^2 + \Gamma_n^2}\;,
\label{eq:decau.1}
\end{equation}
where $\Gamma_n=\Lambda_p \hbar /(2T_p(\epsilon_n))$.

This spectral function tell us that it is possible to construct a
non-stationary solution $\psi_n(r)$ to the Schroedinger equation in the
vicinity of orbit $p$. This solution, let us call it
``scar''\cite{Heller-Tomsovic-93},  would decay to ``flat'' states with the
rate $\Gamma_n/\hbar$. The matrix elements of the decay process $V_{jn}$ are
not known, but it is clear that
\begin{equation}
    \Gamma_n = \pi \sum_j |V_{jn}|^2 A_{n}(\varepsilon_j)\;,
\label{eq:decay.5}
\end{equation}
where $\varepsilon_j$ are energies of the exact wave functions. Fortunately
we will not need explicit values of $V_{jn}$ in further calculations.

In what follows let us assume that electrons fill the system up to the Fermi
energy
at zero temperature. Let $n=d$ is the first ``scar'' state below the Fermi
energy $E_F$.  The Hamiltonian for states in the vicinity of the Fermi
level is
\begin{eqnarray}
   \hat {\cal H} &=&  \sum_\sigma\biggl\{
      \sum_{j, \varepsilon_j > D_d}^{\varepsilon_j < D_{d+1} }
      \left[ \varepsilon_j a_{j\sigma}^\dagger a_{j\sigma}
      + V_{jd} a_{j\sigma}^\dagger a_{d\sigma} + \text{h.c.}
   \right]
\nonumber\\
   &+ & \epsilon_d a_{d\sigma}^\dagger a_{d\sigma}\biggr\}\;,
\label{eq:decay.6a}
\\
   D_d &=& (\epsilon_{d}+\epsilon_{d-1}) /2\;,
\label{eq:decay.6}
\end{eqnarray}
and formulation of the Anderson impurity model\cite{Anderson-may61} is
accomplished by adding the correlation energy
\begin{equation}
   \hat{\cal H}_{\text{corr}} = U \hat n_{d\uparrow}\hat n_{d\downarrow}
\;,
\label{eq:decay.7}
\end{equation}
with
\begin{equation}
   U = \int d\vec r_1 d\vec r_2
   |\psi_{d}(\vec r_1)|^2 |\psi_{d}(\vec r_2)|^2
   {e^2\over |\vec r_1 - \vec r_2|}\;.
\label{eq:decay.8}
\end{equation}
It should be much larger than the charging energy of the dot, $U_j$, given
in general by
\begin{equation}
   U_j = \int d\vec r_1 d\vec r_2
   |\psi_{j}(\vec r_1)|^2 |\psi_{j}(\vec r_2)|^2
   {e^2\over |\vec r_1 - \vec r_2|}\;.
\label{eq:decay.8}
\end{equation}
and almost the same for all flat states $\psi_{j}$. 

The contribution of $p$ to the Green function of the non-interacting
system has the form $G(r,r)\propto e^{iW(x)y^2/\hbar}$, where $\vec r =
(x,y)$,  $x$ is the
coordinate along the orbit $p$, the $y$ axis is perpendicular to the orbit
$p$ at the point
$x$, and $W(x)$ is connected with the second derivatives of the action.
Then, the width of the ``scar'' state is estimated as $\sim \sqrt{\hbar/W}$.
It is clear that $W(x)\sim k/l_p$, where $k=\sqrt{2m\epsilon_d}/\hbar$, and
$m$ is the 
mass of the particle, see for example the analysis of chaotic 
billiards\cite{Bogomolny-88}.  The integral Eq.~(\ref{eq:decay.8}) for the 
wave function concentrated in the rectangle of the width $\sim\sqrt{l_p/k}$
and 
length $l_p$ gives $U\sim {e^2\over l_p} (\log{\sqrt{k l_p}}-1)$. For
experiment of 
Sivan {\em et al}\cite{Sivan-aug96} $k l_p\sim 100$ and the correlation
energy 
$U$ is large.

To summarize the section both a quantum dot and a point contact can have
virtual quasi one-dimensional 
state at the energy $\epsilon_d$; the charging energy of the state is $U$.
This state state is well coupled to a 
finite number of
``flat'' chaotic states in certain energy interval. So we have all
ingredients of the Anderson 
impurity model. It allows to compute valence of the impurity state and other
thermodynamic properties.
Our primary interest is the compressibility of the electron system at
formation/disappearance of magnetic moment.
This regime is described by the almost symmetric limit of the model, see the
next section.

\end{multicols}
\widetext
\begin{figure*}
   \unitlength=1mm
   \linethickness{0.4pt}
   \begin{picture}(140.00,100.00)
   \put(0.00,40.00){\vector(1,0){140.00}}
   \put(70.00,0.00){\vector(0,1){100.00}}
   \put(72.00,95.00){\makebox(0,0)[lc]{$x(\lambda)$}}
   \put(135.00,38.00){\makebox(0,0)[ct]{$\lambda$}}
   \put(70.00,80.00){\line(-1,0){70.00}}
   \put(72.00,80.00){\makebox(0,0)[lc]{$\epsilon_d+U/2$}}
   \bezier{696}(0.00,78.00)(103.00,58.00)(140.00,0.00)
   \put(70.00,76.00){\line(-1,0){61.00}}
   \put(9.00,76.00){\line(0,-1){36.00}}
   \put(9.00,38.00){\makebox(0,0)[ct]{$\lambda_0$}}
   \put(70.00,60.00){\line(-1,0){8.00}}
   \put(62.00,60.00){\line(0,-1){20.00}}
   \put(62.00,38.00){\makebox(0,0)[ct]{$\lambda_1$}}
   \put(72.00,68.00){\makebox(0,0)[lc]{$\Biggr\}\delta\mu$}}
   \put(110.98,35.99){\line(0,-1){3.98}}
   \put(111.00,40.00){\line(0,-1){8.00}}
   \put(111.00,32.00){\line(-1,0){41.00}}
   \put(111.00,42.00){\makebox(0,0)[cb]{$U/8\Gamma-\Gamma/2U$}}
   \put(72.00,30.00){\makebox(0,0)[lt]{$\epsilon_d$}}
   \put(128.00,40.00){\line(0,-1){24.00}}
   \put(128.00,16.00){\line(-1,0){58.00}}
   \put(72.00,14.00){\makebox(0,0)[lt]{$D$}}
   \put(128.00,42.00){\makebox(0,0)[cb]{$\lambda_M$}}
   \end{picture}
   \caption{ The function $x(\lambda)$ as it appears in Bethe equations,
Eq.~\ref{eq:model.3}. 
             The rapidities $\lambda_\alpha$ lie in the interval
$[\lambda_0,\lambda_M]$, corresponding
             energies lie between band bottom $D$ and chemical potential
$\mu\approx x(\lambda_0)$. 
             In the symmetric limit $\mu\sim\epsilon_d+U/2$ and in the
assymetric limit $\mu\sim \epsilon_d$,
             where $\epsilon_d$ is the energy of the virtual state, and $U$
is the charging energy.
}
\label{fig:xotlambd} \end{figure*}
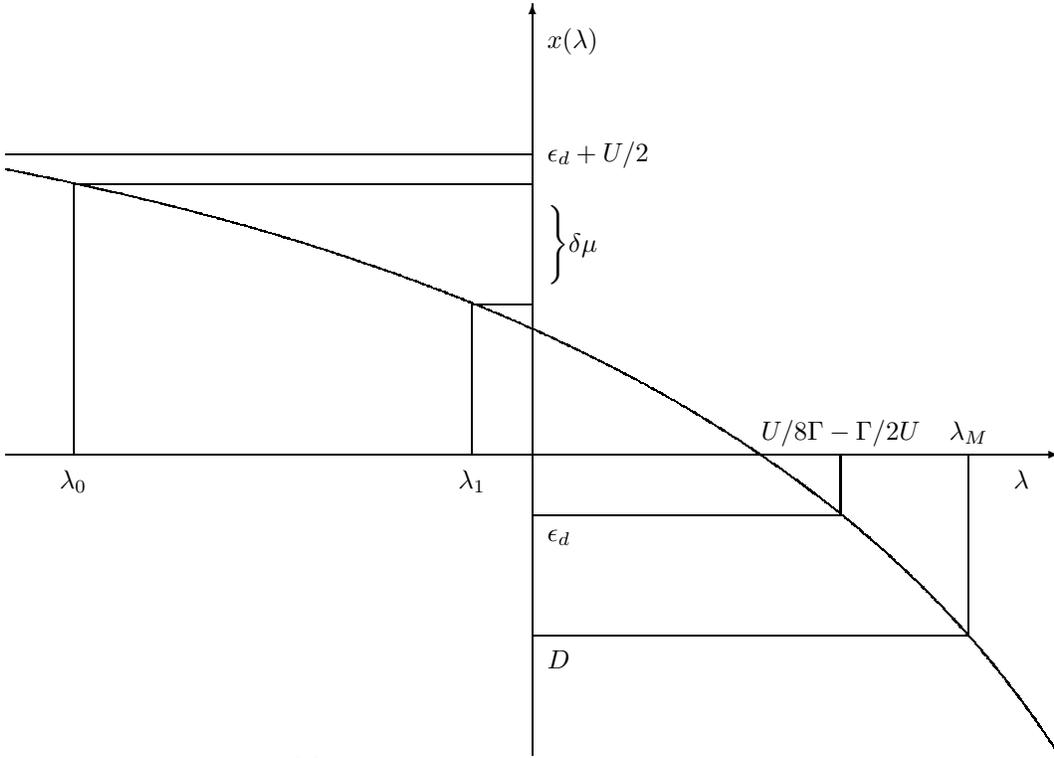

\begin{multicols}{2}
\narrowtext


\section{The level spacing in presence of the Kondo state.}

The purpose of this section is to compute the ground state energy $E_N$ of
$N$-particle 
chaotic system in presence of the Kondo interaction with
quasi-onedimensional state.
Then we will compute the chemical potential $\mu_N=E_{N+1}-E_N$ and the
inverse compressibility 
(the level spacing) of whole system $\Delta_\ast=\mu_{N}-\mu_{N-1}$. This
quantity
is interesting in vicinity of the symmetric regime, when spin is just
disappears and 
we expect strong fluctuations.

Wiegmann have  solved this problem in one dimension\cite{Wiegmann-nov80}. 
However dimensionality of the system is not important since 
$j$-states do not interact with themselves. In order to apply the Wiegmann
solution
we need to postulate the constant energy spacing $2\Delta$ between
$j$-states.
This approximation is not bad in chaotic systems due to level repulsion.

At the zero magnetic field the ground state is described by the set of
rapidities $\{\lambda_\alpha\}_{\alpha = 0}^M$, where $M+1$ is number of
down
spins. The rapidities satisfy
\begin{eqnarray}
  &&
  J_\alpha + {1\over \pi } \sum_{\beta = 0}^M
  \text{arctan}(\lambda_\alpha-\lambda_\beta) 
   = -{1\over \Delta}[ x(\lambda_\alpha) + O(\Delta) ] 
\label{eq:model.2}
\\
  && x(\lambda) = \epsilon_d + U/2  -
  (U\Gamma)^{1/2}(\lambda + (\lambda^2 + 1/4)^{1/2})^{1/2}\;,
\label{eq:model.3}
\end{eqnarray}
$\Gamma = |V|^2$ is also constant. We keep here terms in the lowest order in
$\Delta$.
The integer numbers $\{J_\alpha\}_{\alpha = 0}^M$ completely describe the
state of the
system with $N=2(M+1)$ particles. The energy of the system is just
\begin{equation}
   E = 2\sum_{\alpha=0}^M x(\lambda_\alpha)\;.
\label{eq:model.4}
\end{equation}
The ground state corresponds to subsequent values of $J_\alpha$, fixed by
the condition
\begin{equation}
   J_\alpha < J_M\;,\quad \lambda_\alpha < \lambda_M\;,\quad
   x(\lambda_\alpha) > x(\lambda_M) = D
\label{eq:model.4a}
\end{equation}
where $D$ is the bottom of the energy band, see Fig.~\ref{fig:xotlambd}.

The ground state is a bit more complicated for odd number of particles, $N =
2M+1$. 
In this case $\alpha = 0\ldots M-1$, and we introduce new rapidity $\tilde
\lambda$:
\begin{eqnarray}
  &&
  J_\alpha +
  {1\over \pi } \sum_{\beta = 0}^{M-1}
  \text{arctan}(\lambda_\alpha-\lambda_\beta) 
  + {1\over \pi}  
  \text{arctan}(2\lambda_\alpha-2\tilde\lambda) 
\nonumber
\\
   && = -{1\over \Delta} x(\lambda_\alpha) 
\label{eq:model.2a} 
\\
  && \tilde x(\tilde \lambda) = \epsilon_d + U/2  -
(2U\Gamma\tilde\lambda)^{1/2} = D \;,
\label{eq:model.3a}
\\
   && E = \tilde x(\tilde \lambda) + 2\sum_{\alpha=0}^{M-1}
x(\lambda_\alpha)\;.
\label{eq:model.4b}
\end{eqnarray}
In order to compute the chemical potential we need both solutions
Eq.~(\ref{eq:model.4}) and Eq.~(\ref{eq:model.4b}), 
because one additional electron changes number of particles from odd to even
or vise versa. Fortunately,
the difference between the odd-electron solution and the even-electron
solution is small, something happens 
at the bottom of the band, and not near the Fermi surface. For this very
reason we will define the chemical
potential as half of energy one needs to add two particles to a system with
even number of particles:
\begin{equation}
   \mu_N \equiv { E_{N+2} - E_{N} \over 2 }\;,\quad
   \Delta_\ast = { \mu_{N} - \mu_{N-2} \over 2 }\;.
\label{eq:model.4c}
\end{equation}
Here $\Delta_\ast$ is the level spacing of interacting system.

When the level spacing of non-interacting system is small
$\Delta\rightarrow0$, one can solve 
Eq.~(\ref{eq:model.2}) by making use of the continuous approximation. It is
justified by the condition
\begin{equation}
   |\lambda_{\alpha+1}-\lambda_\alpha| \ll 1\;\;\;\;\forall \alpha\;.
\label{eq:model.5}
\end{equation}
So we replace sum Eq.~(\ref{eq:model.2}) by integral
\begin{equation}
  J_\alpha + {1\over \Delta } \int_{\lambda_0}^{\lambda_M}
  \text{arctan}(\lambda_\alpha-\lambda')\sigma(\lambda')d\lambda' 
   = -{1\over \Delta} x(\lambda_\alpha)\;,
\label{eq:model.2c}
\end{equation}
where $\sigma(\lambda) \sim {\Delta /\pi \over \lambda_{\alpha+1} -
\lambda_\alpha}$. 
Taking derivative with respect to $\lambda_\alpha$ we arrive at the integral
equation
\begin{equation}
   \sigma_\lambda + {1\over \pi}\int_{\lambda_0}^{\lambda_M}
   {\sigma_{\lambda'} d\lambda'
       \over
   1 + (\lambda - \lambda')^2}
   = - {1\over \pi}{dx \over d\lambda}
\label{eq:model.6}
\end{equation}
which can be formally solved in whole range $\lambda\in
]-\infty,\lambda_M]$.
For large $\lambda_M$ we can write equation with $\alpha=M$ as $J_M + M/2
\approx -D/\Delta$, then
all $J_\alpha$s become fixed $J_\alpha = -D/\Delta - 3M/2 + \alpha$.
Integrating Eq.~(\ref{eq:model.6}) 
from $-\infty$ to $\lambda_\alpha$ (again $\lambda_M$ is large)
we get equation for $\lambda_\alpha$:
\begin{equation}
    \pi\int_{-\infty}^{\lambda_\alpha}
    \tilde\sigma_{\lambda} d\lambda =
    \epsilon_d + U/2 - D - 2M\Delta +\alpha\Delta\;.
\label{eq:model.7}
\end{equation}
where we made use of $x(-\infty)=\epsilon_d+U/2$.

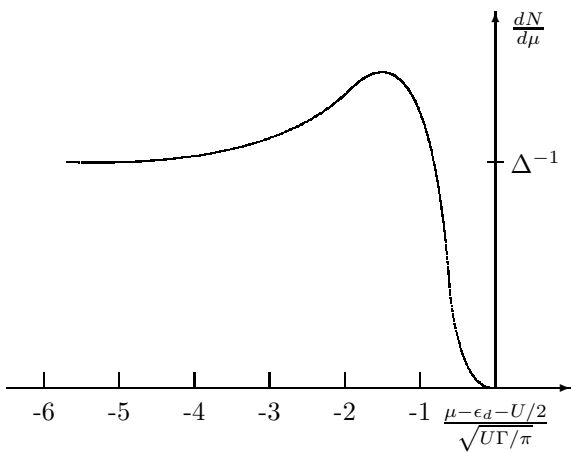
\begin{figure}
\unitlength 1mm
\linethickness{0.4pt}
\begin{picture}(80.00,60.00)
\put(5.00,10.00){\vector(1,0){75.00}}
\put(70.00,10.00){\vector(0,1){50.00}}
\bezier{68}(69.99,10.02)(65.72,9.26)(63.96,21.94)
\bezier{212}(63.96,21.94)(61.33,59.97)(50.78,49.81)
\bezier{168}(50.78,49.81)(40.36,39.14)(13.00,40.02)
\put(69.00,40.00){\line(1,0){2.00}}
\put(60.00,12.00){\line(0,-1){2.00}}
\put(50.00,12.00){\line(0,-1){2.00}}
\put(40.00,12.00){\line(0,-1){2.00}}
\put(30.00,12.00){\line(0,-1){2.00}}
\put(20.00,12.00){\line(0,-1){2.00}}
\put(10.00,12.00){\line(0,-1){2.00}}
\put(72.00,60.00){\makebox(0,0)[lt]{${dN \over d\mu}$}}
\put(72.00,40.00){\makebox(0,0)[lc]{$\Delta^{-1}$}}
\put(60.00,8.00){\makebox(0,0)[ct]{-1}}
\put(50.00,8.00){\makebox(0,0)[ct]{-2}}
\put(40.00,8.00){\makebox(0,0)[ct]{-3}}
\put(30.00,8.00){\makebox(0,0)[ct]{-4}}
\put(20.00,8.00){\makebox(0,0)[ct]{-5}}
\put(10.00,8.00){\makebox(0,0)[ct]{-6}}
\put(70.00,8.00){\makebox(0,0)[ct]
{$\mu - \epsilon_d - U/2 \over \sqrt{U\Gamma /\pi} $}}
\end{picture}
\caption{The compressibility of electron gas in presence of the Anderson
impurity. Electron system becomes 
incompressible in the mixed-valence regime.  }
\label{fig:compr}
\end{figure}

The general (Wienner-Hopf) solution of Eq.~(\ref{eq:model.6}) is available
in the 
Wiegmann paper\cite{Wiegmann-sep82}. The particular case $\lambda_0=-\infty$
and $\lambda_M=\infty$ can be solved 
by Fourier transform method and the result is
\begin{equation}
   \sigma_\lambda
    = \int_{-\infty}^\infty {dk\over 2\pi}
    { 1\over \cosh (\pi\lambda - \pi k^2/(2U\Gamma))}
\label{eq:model.6b}
\end{equation}
it changes a little when $\lambda_0, \lambda_M$ are finite but large. 
For large negative $\lambda$ the solution behaves exponentially
\begin{equation}
   \sigma_\lambda
    \approx \int_{-\infty}^\infty {dk\over 2\pi} 2 e^{\pi\lambda - \pi
k^2/(2U\Gamma)}
    = \sqrt{ 2 U\Gamma } e^{\pi\lambda}
\label{eq:model.6a}
\end{equation}
Then we can compute the lowest rapidity $\lambda_0$ from
Eq.~(\ref{eq:model.7})
\begin{equation}
   \lambda_0 = {1\over \pi}
   \log{\epsilon_d + U/2 - D - 2 M \Delta \over \sqrt{2U\Gamma} }
\label{eq:model.8}
\end{equation}
and indeed it goes to $-\infty$ in the symmetric limit of the Anderson model

$\epsilon_d + U/2 - D - 2M\Delta \rightarrow 0$, 
( it is enough to have $|\epsilon_d + U/2  + D - 2M\Delta |\ll
\sqrt{U\Gamma}$). 

Solution Eq.~(\ref{eq:model.6b}) works for any $\lambda_0$ if
$\lambda<\lambda_0$. Then we combine
Eq.~(\ref{eq:model.6b}) with Eq.~(\ref{eq:model.3}) and obtain
compressibility
\begin{equation}
   {dN \over d\mu} =
    {1\over \Delta}
    \int_{-\infty}^\infty {dk\over 2\pi}
    { -\tilde\mu - \pi^2/4 \tilde \mu^3 \over \cosh( - ( \pi^2/4 - \tilde
\mu^4 ) / 2\tilde \mu^2 - k^2/2) }
\label{eq:model.8a}
\end{equation}
where $\tilde \mu = (\mu - \epsilon_d - U/2)/\sqrt{U\Gamma/\pi}$. The
compressibility is zero at symmetric
limit, see Fig.~\ref{fig:compr}.

In the vicinity of symmetric limit one has simple expressions for chemical
potential and level
spacing
\begin{eqnarray}
   \mu_{N} &=& x(\lambda_0) 
\nonumber\\
   &=& \epsilon_d + U/2 - 
   \sqrt{\pi U\Gamma/8\over \log (\sqrt{2U\Gamma}/A_N) } \;,
\nonumber
\\
   \Delta_\ast &=& {\partial\mu\over \partial N} = {\Delta \over A_N}
   \sqrt{\pi U\Gamma/32 \over \log^3(\sqrt{2U\Gamma}/A_N)}\;,
\label{eq:model.8a}
\\ A_N &=& \epsilon_d + U/2 - D - (N-2) \Delta \;.
\end{eqnarray}
The level spacing $\Delta_\ast$ diverges for any fixed $\Delta$ in the
vicinity of the symmetric limit.
This result indicates that system is unstable near $\mu=\epsilon_d + U/2$,
and experimentally one observes
kinks of level spacing. 

The kink of the level spacing means that the system becomes 
non-interacting, the chemical potential get its $U=0$ value
\begin{equation}
    \mu_N|_{U=0} = N\Delta + D\;,
\label{model.res}
\end{equation}
and the observed kink is
\begin{eqnarray}
   \Delta_\ast &=& \mu_N|_{U=0} - \mu_{N-2}\;,
\nonumber
\\
   &=& 
   \sqrt{\pi U\Gamma/8\over \log (\sqrt{2U\Gamma}/A_{N-2} ) } - A_N\;.
\label{eq:model.res2}
\end{eqnarray}
The maximal possible kink of the level spacing is
\begin{equation}
    \Delta_{\text{max}} = \sqrt{\pi U\Gamma/4\over \log (U\Gamma/2\Delta) }
\gg \Delta\;.
\label{eq:model.res}
\end{equation}
It goes to zero in thermodynamic limit $\Delta\rightarrow 0$, but remains
large. 
We will derive Eqs.~(\ref{eq:model.res},{eq:model.res2}) in the rest of the
section.

{\em The problem } is that the condition Eq.~(\ref{eq:model.5}) is broken
for the lowest $\lambda_\alpha$
near the symmetric limit. Since for $\lambda_0\rightarrow-\infty$ the
density of the rapidities goes to zero exponentially, 
so  $|\lambda_1-\lambda_0|\propto e^{\pi|\lambda_0|}\Delta/\sqrt{U\Gamma}$
can be very large for any fixed $\Delta$.
The main idea of the present paper is to introduce one rapidity $\lambda_0$
separated from all the rest $\lambda_1\ldots\lambda_M$. We should assume 
$|\lambda_0-\lambda_1|\gg1$ in Eq.~(\ref{eq:model.2c}) with $\alpha=0$:
\begin{equation}
   J_0  - {M\over 2} - 
   {1\over \Delta}\int_{\lambda_1}^{\lambda_{M}} 
   {\sigma_{\lambda} d\lambda \over \lambda_0 - \lambda}
   = - {1\over \Delta}
   (\epsilon_d + U/2 - \sqrt{U\Gamma\over 8|\lambda_0|})
\label{eq:model.9}
\end{equation}
where we expanded arctan(). The integral equation for region of "dense"
rapidities becomes
\begin{equation}
   \sigma_\lambda + {\Delta/\pi \over 1 + (\lambda - \lambda_0)^2}
   {1\over \pi}\int_{\lambda_1}^{\lambda_M}
   {\sigma_{\lambda'} d\lambda'
       \over
   1 + (\lambda - \lambda')^2}
   = - {1\over \pi}{dx \over d\lambda}
\label{eq:model.11}
\end{equation}
In the lowest order in $1/ \lambda_0$ we neglect the integral in
Eq.~(\ref{eq:model.9}) and obtain 
$x(\lambda_0)/\Delta=M/2 - J_0 = 2M + D$ that is equivalent to
Eq.~(\ref{eq:model.res}). Lowest
rapidities are
\begin{equation}
  \lambda_0 \approx -{ U\Gamma/8 A_N^2}\;,\quad
   \lambda_1 = {1\over \pi} \log\left(A_{N-2} / \sqrt{2U\Gamma} \right)\;,
\label{eq:model.12}
\end{equation}
where $\lambda_1$ goes to $-\infty$ much slowly than $\lambda_0$. 
Within the approximation made here $\lambda_1$ of system of 
$N+2$ particles coincides with $\lambda_0$ of system of $N$ particles. Then
the obtained pseudogap 
Eq.~(\ref{eq:model.res}) is just $x(\lambda_0)-x(\lambda_1)$.

In conclusion of the section we have found the anomalously large spacing
between
energy levels of electron gas interacting with impurity. This is a 
finite size effect, which requires special treatment.
The pseudogap appears at the Fermi energy when 
the model approaches complete particle-hole symmetry,
$\mu \approx \epsilon_d+U/2$.


\section{ Experimental data and conclusions.}

Let us consider experiment, where one fills the system by particles. The
Fermi level goes up $\mu(N)\sim N\Delta + D$, where $N$ is the
number of particles inside the system, and $D$ is the band offset.
At the value $N=N_d$ given by equation
\begin{equation}
   \mu(N_d)  =  \epsilon_d + U/2
\label{eq:model.1}
\end{equation}
we arrive at the symmetric Anderson model. Just below this value of $N$ the
Fermi energy
makes a large jump given by Eq.~(\ref{eq:model.res2}). Later we will measure
the $\Delta_\ast$
in terms of mean level spacing
\begin{equation}
   \delta_\ast = \Delta_\ast/\Delta \approx \sqrt{(\pi/4)x/\log x}\;,
   \quad x = U\Gamma/2\Delta^2 \gg 1\;.
\label{eq:model.3}
\end{equation}
The Fermi energy will jump again near $N=N_{d+1}$. The distance between
jumps is 
$ N_{d+1}-N_d \approx (\epsilon_{d+1}-\epsilon_d) / \Delta $ is inverse
proportional to the length of
the orbit $l_p$
\begin{equation}
   \delta N = N_{d+1} - N_{d} \approx {2L\over l_p}
   \sqrt{2\pi N_{d+1}}\;.
\label{eq:model.3}
\end{equation}
This formula was derived for $\Delta = \pi \hbar^2/(m A)$, where $L$ is size
of the
quantum dot, $A\approx L^2$ is its area and $\delta N \ll N_{d+1}$. 

The experimental data from few references is summarized in
Table~\ref{tab:table1}. The first block contains 
data taken from cited papers, then we provide estimation of $k_F$, $\Delta$,
$\Gamma$, $U$, $\delta N$, and 
$\delta_\ast$. Last block in Table~\ref{tab:table1} contains $\delta N$ and
$\delta_\ast$ taken from 
charts of $E_{N+1}-E_{N}$ vs $N$ reported in cited papers.

Typically $l_p\sim 2L$ and we will use $\delta N = \sqrt{2\pi N_{d+1}}$. The
Coulomb energy of the 
virtual state is $ U \approx E_C \log k_F l_p $ where $E_C$ is the charging
energy of the dot, 
and $k_F$ is wave number on Fermi surface. The coupling $\Gamma$ is of the
order of level spacing of 
virtual state, at least it larger than level spacing of host system: 
$\Delta \lesssim \Gamma\lesssim \epsilon_{d+1}-\epsilon_d$. For this reason
Table~\ref{tab:table1} contains
two values for $\Gamma$ - minimal and maximal estimates. Last computed value
is $\delta_\ast$, which eventually
have minimal and maximal estimates derived from Eq.~(\ref{eq:model.3}).
We see that our theory works well for clean
samples\cite{Sivan-aug96,Simmel-apr97,Patel-may98} and it is not applicable
for disordered quantum dots\cite{Zhitenev-sep97}. The virtual state is
absent in perfectly rectangular quantum 
dot\cite{Lucher-mar2001} (without shape deformations) as well.

The conductance of a quantum point contact (QPC) with the Kondo state inside
has been computed recently\cite{Meir-2002} and
the model explained all the experimental data.  Existence of the Kondo state
in a QPC is very plausible,
and therefore we expect to see Kondo state inside quantum dots too. The
present theory should not be confused with Kondo
effect due to charging of entire dot\cite{Kouwenhoven-Glazman}. However the
intersting feature of both types of experiments
(Kondo state on ``scar'' and Kondo state inside entire dot) is that the gate
voltage allows to scan all regimes
of Anderson model\cite{Goldhaber-Gordon-1998}.

To summarize, inhomogeneities of chaotic wave functions can be gathered
together into an additional state of small size. In this way one arrives at
Anderson's impurity model with finite number of free electron states. At
certain value of the chemical potential the model has complete particle-hole
symmetry. The chemical potential does not approach this value
gradually, but rather irregularly. The energy scale 
of this effect is a new combination of the parameters of  Anderson's model
and the mean level spacing. This scenario is a possible explanation of the 
irregularities observed experimentally in quantum dots.


\begin{table}
\caption{
Measured data / Intermediate calculations/ Observed values of $\delta N$ and
$\delta_\ast$.
energies are given in meV, 
length are given in $\mu$m and wavenumbers in $\mu$m$^{-1}$.
}
\begin{center}
\begin{tabular}{c|c|c|c|c}
Ref. &   \onlinecite{Sivan-aug96} & \onlinecite{Simmel-apr97} &
\onlinecite{Zhitenev-sep97} & \onlinecite{Patel-may98}$^\ddagger$ \\
\hline
$N$ &                         100 &                      250  &
100  &                      340  \\
$A$ &                        0.15 &                      0.07 &
0.13 &                      0.17 \\
$L$ &                         0.5 &                      0.3  &
1.2$^\dagger$ &                       0.9 \\
$E_C$ &                       0.6 &                       2   &
1    &                      0.59 \\
\hline
$k$ &                          65 &                      150  &
83  &                      110  \\
$\Delta$ &                   0.02 &                      0.05 &
0.04 &                      0.02 \\
$U$ &                         2.0 &                       8   &
4.6  &                      3    \\
$\Gamma$    &          0.02 - 0.5 &                0.05 - 2.0 &
0.04 - 1.0 &               0.02 - 1.0  \\
$\delta N $ &                  25 &                       40  &
25  &                      46   \\
$\delta_\ast $ &           3 - 10 &                   4 - 16  &
3 - 7 &                    3 - 13 \\
\hline
$\delta N $&                   15 &                       22  &
5   &                       -   \\
$\delta_\ast$                & 10 &                       6-8 &
2   &                        6  \\
\end{tabular}\\
\end{center}
$\dagger$ -- circumference
$\ddagger$ -- the sample \# 1
\label{tab:table1}
\end{table}

\end{multicols}
\end{document}